\documentclass[fleqn,usenatbib,letters]{mnras}

\usepackage{mathptmx}

\usepackage[T1]{fontenc}
\usepackage{ae,aecompl}

\usepackage{graphicx}	
\usepackage{amsmath}	
\usepackage{amssymb}	
\usepackage{wasysym}

\usepackage{multirow}
\usepackage{arydshln}
\usepackage{wasysym}	

\usepackage{upgreek}

\defcitealias{Lecav2013}{L13}

\title[Lightning on HAT-P-11b?]{Is lightning a possible source of the radio emission on HAT-P-11b?}

\author[Hodos\'{a}n et al.]{G. Hodos\'an$^{1}$\thanks{E-mail:
gh53@st-andrews.ac.uk}, P. B. Rimmer$^{1}$ and Ch. Helling$^{1}$\\
$^{1}$SUPA, School of Physics and Astronomy, University of St Andrews, St Andrews KY16 9SS, UK \\
}

\date{Accepted 2016 April 21; Received 2016 April 20; in original form 2016 March 17}

\pubyear{2016}

\begin{document}
\label{firstpage}
\pagerange{\pageref{firstpage}--\pageref{lastpage}}
\maketitle

\begin{abstract}
Lightning induced radio emission has been observed on Solar system planets. There have been many attempts to observe exoplanets in the radio wavelength, however, no unequivocal detection has been reported. Lecavelier des Etangs et al. carried out radio transit observations of the exoplanet HAT-P-11b, and suggested that a small part of the radio flux can be attributed to the planet. Here, we assume that this signal is real, and study if this radio emission could be caused by lightning with similar energetic properties like in the Solar system. We find that a lightning storm with $3.8 \times 10^6$ times larger flash densities than the Earth-storms with the largest lightning activity is needed to produce the observed signal from HAT-P-11b. The optical emission of such thunderstorm would be comparable to that of the host star. We show that HCN produced by lightning chemistry is observable $2-3$ yr after the storm, which produces signatures in the \textit{L} ($3.0-4.0 \mu$m) and \textit{N} ($7.5-14.5 \mu$m) infrared bands. We conclude that it is unlikely that the observed radio signal was produced by lightning, however, future, combined radio and infrared observations may lead to lightning detection on planets outside the Solar system.
\end{abstract}




\section{Introduction}
\label{sec:intro}

The discovery of the first exoplanet around a neutron star \citep{Wolszczan1992} and then around a Sun-like star \citep{Mayor1995} opened the gates to a new astronomical field concerning extrasolar planetary systems. By March 2016, there have been 2087 exoplanets discovered \citep[including 1297 transiting and 65 directly imaged planets, the most suitable types for lightning-hunting,][]{Hodosan2016a}\footnote{http://exoplanet.eu/; 2016.03.08}. This large number allows us to focus on the more detailed characterization of the different types of planets, including atmospheric chemistry and internal composition of the planetary bodies. Recently, radio observations have opened new paths to study properties of extrasolar objects, such as brown dwarfs \citep[e.g.][]{Williams2015}, which are only a step away from giant gas planet detections in the radio wavelengths. Electron cyclotron maser emission has been suggested to be one possible mechanism for the radio emission \citep{Griessmeier2007}. A campaign to observe this emission has been started \citep[e.g.][and ref. therein]{Lecav2013}; however, no electron cyclotron maser emission from exoplanets has been conclusively detected.

\citet[][]{Lecav2013} \citepalias[hereafter][]{Lecav2013} presented a tentative detection of a radio signal from the exoplanet HAT-P-11b. This planet is estimated to have a radius of 4.7 R$_{\oplus}$ (R$_{\oplus}$: Earth radius), a mass of 26 M$_{\oplus}$ (M$_{\oplus}$: Earth mass), and is at a distance of $\sim 0.053$ au from its host star \citep{Bakos2010, Lopez2014}. In 2009, \citetalias{Lecav2013} observed a radio signal at 150 MHz with an average flux of 3.87 mJy that vanished when the planet passed behind its host star. They re-observed the planet with the same instruments in 2010, but no signal was detected this time. Assuming that the 150 MHz signal from 2009 is real and from the exoplanet, the non-detection in 2010 suggests that it is produced by a transient phenomenon. \citetalias{Lecav2013} suggested that the obtained radio signal is the result of interactions between the planetary magnetic field and stellar coronal mass ejections or stellar magnetic field. If the radio signal is real, it is unlikely to be due to cyclotron maser emission, because this type of emission is generally polarized \citep{Weibel1959, Vorgul2016}, and fig. 2 in \citetalias{Lecav2013} shows a non-detection of polarization in the data. If the mJy radio emission were caused by cyclotron maser emission, a large planetary magnetic field of 50 G would be required \citepalias{Lecav2013}.

Lightning is a transient phenomenon of several known cloudy atmospheres, with flash rates varying greatly. Lightning induced radio emission has been observed in the Solar system on Earth, most probably on Venus \citep[e.g.][]{Russell2008}, and on the giant planets Jupiter \citep[e.g.][]{Gurnett1979}, Saturn \citep[e.g.][]{Fischer2006,Fischer2007}, Uranus \citep{Zarka1986} and Neptune \citep{Aplin2013}. \citet{Bailey2014} suggested that present day radio observations of brown dwarfs may contain hints to the presence of lightning in these planetary-like atmospheres. Lightning emission is not polarized \citep{Shao2002}. Therefore, based on the non-detection of polarization in combination with the possible transient nature of the observed radio emission, we tentatively hypothesize that the emission on HAT-P-11b is caused by lightning discharges.

HAT-P-11b is much closer to its host star than the Solar system planets with lightning, resulting in a stronger irradiation from the star. 3D simulations of irradiated giant gas planets have demonstrated that a very strong circulation of the atmosphere results from the high irradiation \citep[e.g.][]{Heng2015}, and it seems reasonable to expect similar effects for Neptune-like planets. Works like \citet{Lee2015} and \citet{Helling2016} further suggest that highly irradiated atmospheres will form clouds in very dynamic environments, which may host high lightning activity, because triboelectric charging in combination with gravitational settling allows lightning discharge processes to occur in extrasolar clouds \citep{Helling2013}. \citet{Fraine2014} took the transmission spectra of HAT-P-11b and interpreted the data with a clear atmosphere model. However, \citet{Line2016} found that in the case of HAT-P-11b, patchy clouds could explain these transmission spectra. HAT-P-11b, orbiting its host star closely, likely has a dynamic atmosphere that will impact the observable cloud distribution. The resulting patchy clouds could focus potential lightning activity to a certain region, maybe at certain times covering a large fraction of the planet. These potentially large, dynamical cloud systems could support the occurrence of high lightning rates in particular regions.

\section{Radio signal strength and lightning frequency}
\label{sec:radio}

In this section we calculate the radiated power spectral density of lightning at 150 MHz, which allows us to estimate the radio flux of one lightning flash at this frequency. We aim to derive a lower limit for the lightning flash density, $\rho_{\rm fl}$ [flashes km$^{-2}$ h$^{-1}$], that would be needed to reproduce the intermittent radio emission from HAT-P-11b.

The power spectral density, $P/\Delta f$ [W Hz$^{-1}$] of lightning radiated at a frequency $f$ [Hz] follows a power law: 

\begin{equation}
\label{eq:1}
\dfrac{P}{\Delta f} = \dfrac{P_0}{\Delta f} \Bigg(\dfrac{f_0}{f}\Bigg)^{\!\! n},
\end{equation}

\noindent where $P_0/\Delta f$ [W Hz$^{-1}$] is the peak power spectral density at a peak frequency of $f_0$ [Hz], and $n$ is the spectral roll-off at higher frequencies \citep{Farrell2007}. The spectral irradiance of a single lightning flash, $I_{\rm \nu, fl}$, is related to $P/\Delta f$: 

\begin{equation}
\label{eq:2b}
I_{\rm \nu, fl} = \frac{(P/\Delta f)}{4 \uppi d^2} \times 10^{26},
\end{equation}

\noindent where $d$ is the distance to the HAT-P-11 system, and 1 W Hz$^{-1}$ m$^{-2} = 10^{26}$ Jy. The observed spectral irradiance, $I_{\rm \nu, obs}$:

\begin{equation}
\label{eq:2}
I_{\rm \nu, obs} = I_{\rm \nu, fl} \frac{\tau_{\rm fl}}{\tau_{\rm obs}} n_{\rm tot,fl},
\end{equation}

\noindent where $\tau_{\rm fl}$ [h] is the characteristic duration of the lightning event, $\tau_{\rm obs}$ [h] is the time over which the observations were taken, and $n_{\rm tot,fl}$ is the total number of flashes. A lightning flash has a much shorter duration than the observation time, therefore it cannot be considered as a continuous source. As a result, the contribution of one lightning flash ($I_{\rm \nu, fl}$) to the observed spectral irradiance ($I_{\rm \nu, obs}$) has to be weighted by its duration time ($\tau_{\rm fl}$) over the observation time ($\tau_{\rm obs}$).

From equation (\ref{eq:2}) we obtain $n_{\rm tot,fl}$, and then calculate the flash density,  $\rho_{\rm fl}$ [flashes km$^{-2}$ h$^{-1}$]:

\begin{equation}
\label{eq:2a}
\rho_{\rm fl} = \frac{n_{\rm tot, fl}}{2\uppi R_p^2 \tau_{\rm obs}},
\end{equation}

\noindent where $R_p$ [km] is the radius of the planet. Equation (\ref{eq:2}) gives the total spectral irradiance resulted from lightning flashes from over the projected disk of the planet ($2\uppi R_p^2$). 

In order to estimate the flash density that would result in a radio signal like the one obtained by \citetalias{Lecav2013}, we assume that lightning on HAT-P-11b has the same energetic properties as lightning on Saturn. \textit{Cassini-RPWS} measured the radiated power spectral density of lightning on Saturn to be $P/\Delta f =$ 50 W Hz$^{-1}$ at $f =$ 10 MHz \citep{Fischer2006, Farrell2007}. We use equation (\ref{eq:1}) and the values observed by the \textit{Cassini} probe for $P/\Delta f$ and $f$ to obtain a peak spectral power density of $P_0/\Delta f = 1.6 \times 10^{12}$ W Hz$^{-1}$, for $f_0 = 10$ kHz and $n = 3.5$, the most favourable values for minimizing flash densities in the case of HAT-P-11b.\footnote{The Earth value $f_0 = 10$ kHz \citep{Rakov2003} and a gentler spectral roll-off, $n = 3.5$ ($n = 4$ for Earth) were used because these values are not known for any other Solar system planet. These values are used for modelling lightning on Jupiter or Saturn \citep[e.g.][]{Farrell2007}.} Next, by applying $P_0/\Delta f$ to equation (\ref{eq:1}) we estimated the radiated power spectral density at the source of a single lightning flash at $f = 150$ MHz, frequency at which the HAT-P-11b radio signal was observed \citepalias{Lecav2013}, to be $P/\Delta f = 3.9 \times 10^{-3}$ W Hz$^{-1}$.

\begin{figure*}
\includegraphics[width=\columnwidth]{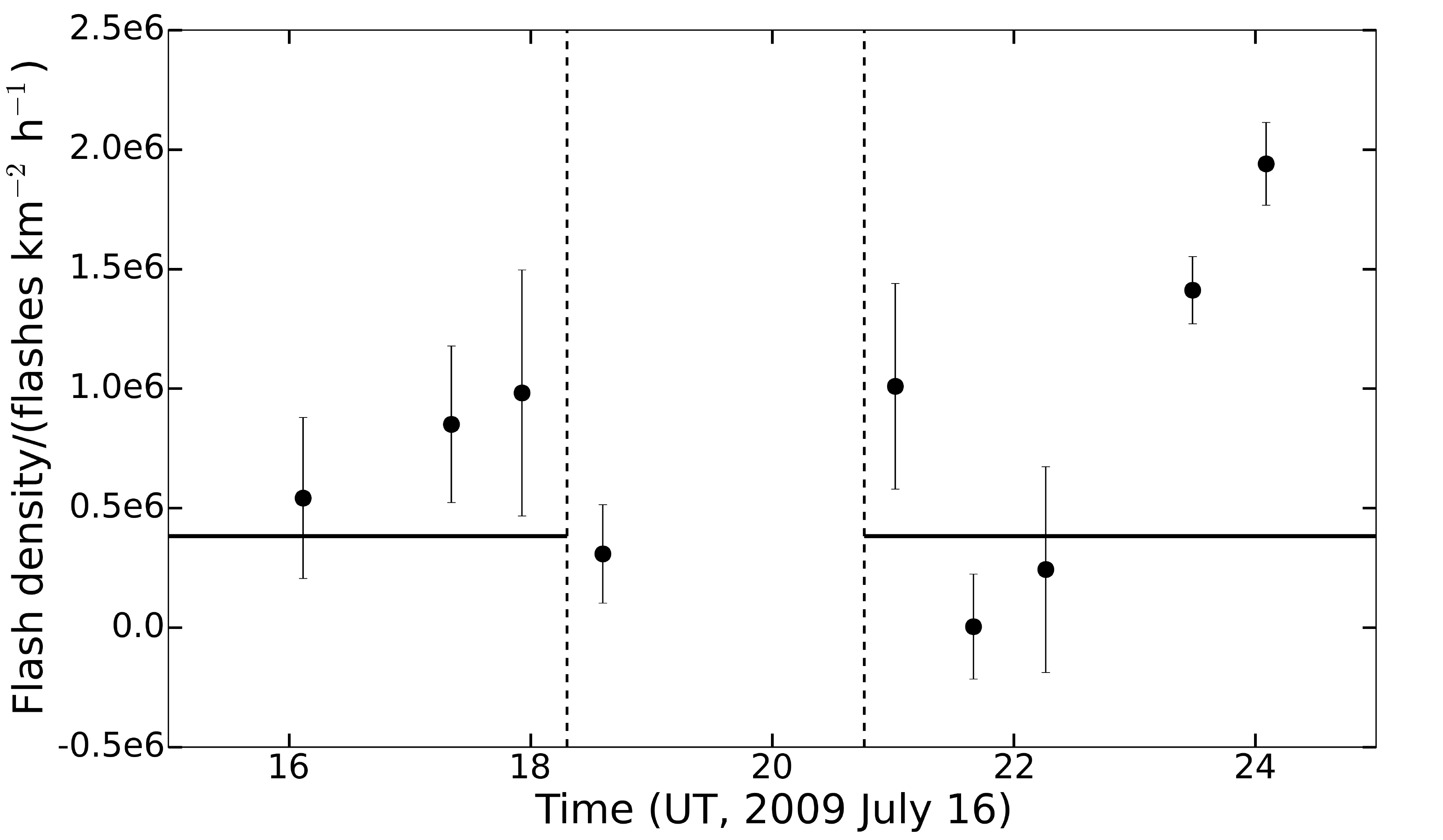}
\includegraphics[width=\columnwidth]{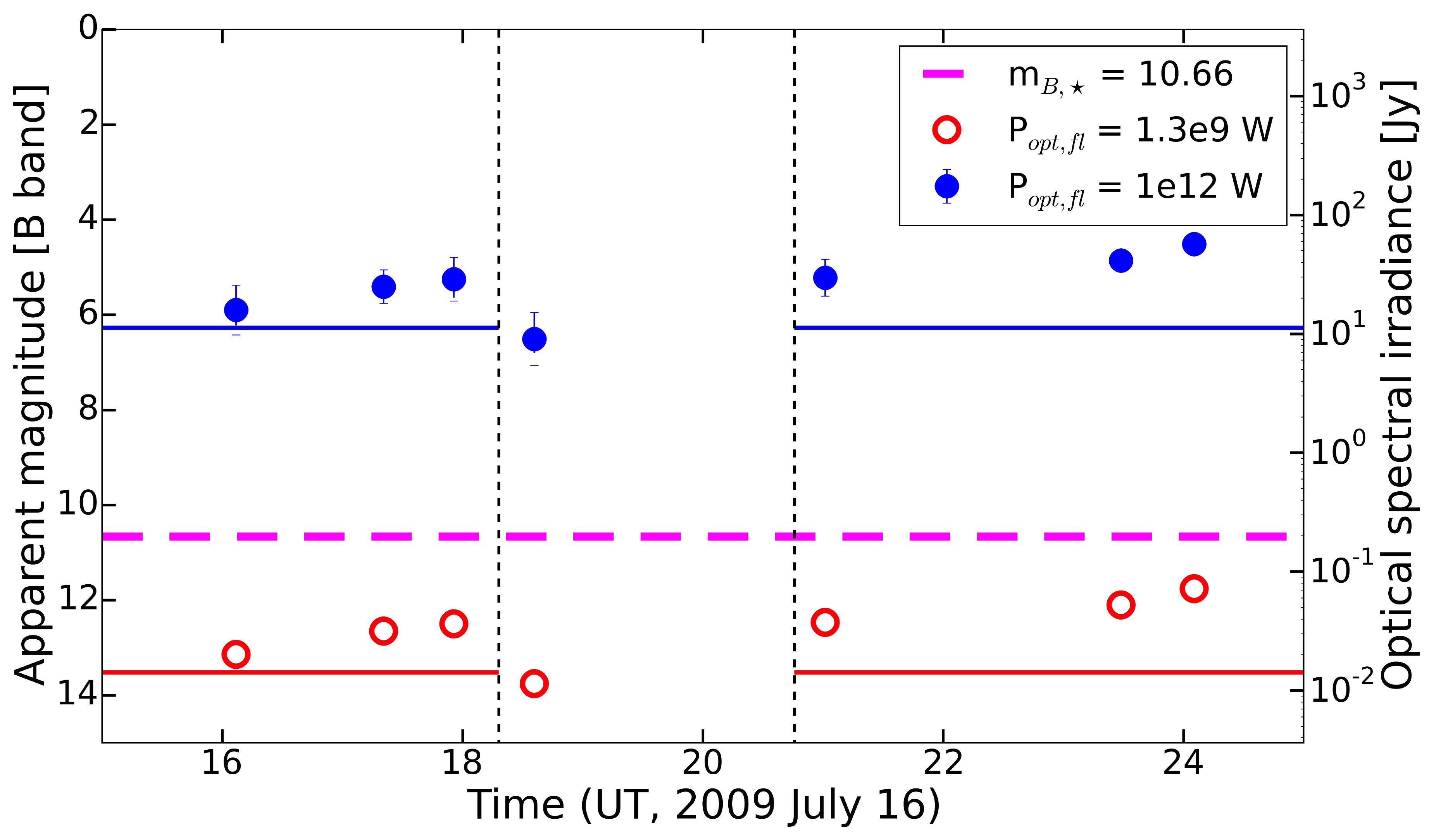}
\caption{Lightning flash densities (left) and apparent magnitude of the lightning flashes (right) that would produce the radio fluxes observed by \citetalias{Lecav2013} for HAT-P-11b (their fig. 2). Horizontal solid lines: average values for the average observed radio flux of 3.87 mJy outside eclipse. Vertical dashed lines: beginning and end of the secondary eclipse of the planet. We show the mean results for the range of observed values per time from \citetalias{Lecav2013}. \textbf{Right:} Results for two different optical powers, Saturnian ($1.3 \times 10^{9}$ W; red) and terrestrial super-bolt ($10^{12}$ W; blue). Magenta dashed line: apparent B magnitude of the host star HAT-P-11.}
\label{fig:rflux}
\end{figure*}

Using equation (\ref{eq:2b}) and the distance of HAT-P-11, $d = 38$ pc, we obtain the spectral irradiance for a single lightning flash to be $I_{\rm \nu, fl} = 2.2 \times 10^{-14}$ Jy. \citetalias{Lecav2013} found the average observed spectral irradiance, $I_{\rm \nu, obs}$, to be  3.87 mJy. Solving equation (\ref{eq:2}) for $n_{\rm tot,fl}$, the total number of lightning flashes needed to explain the observed spectral irradiance, with an average event duration, $\tau_{\rm fl} = 0.3$ s on Saturn \citep[the largest event duration according to][]{Zarka2004}, we obtain a value of $n_{\rm tot,fl} \approx 1.3 \times 10^{15}$ flashes. The integration time for a single data point in \citetalias{Lecav2013} (their fig. 2) is $\tau_{\rm obs} = 36$ min, and the radius of HAT-P-11b is $R_p \approx 0.4$ R$_{\rm J}$ \citep[R$_{\rm J}$: Jupiter radius;][]{Bakos2010}. Substituting these values and the derived $n_{\rm tot,fl}$ into equation (\ref{eq:2a}) we obtain a flash density of $\rho_{\rm fl} \approx 3.8 \times 10^5$ flashes km$^{-2}$ h$^{-1}$ Figure \ref{fig:rflux} (left) shows the flash densities that would be needed to produce a radio flux comparable to observations in \citetalias{Lecav2013} (their fig. 2), assuming that the flux is from the planet and is entirely produced by lightning activity.

\citet{Fischer2011b} analysed the SED (Saturnian Electrostatic Discharges) occurrence of a giant storm that occurred on Saturn in 2010/2011. They found an SED rate of 10 s$^{-1}$, which is 36000 SED h$^{-1}$. This is the largest rate observed on Saturn. Since no other storms were observed during this period \citep{Dyudina2013}, we apply this flash rate (assuming that one SED originates from one flash) for the whole surface of the planet. This results in a flash density of $8.4 \times 10^{-7}$ flashes km$^{-2}$ h$^{-1}$ for Saturn \citep[e.g][]{Hodosan2016a}. The observed signal on HAT-P-11b would require a storm with flash densities $\sim 4.5 \times 10^{11}$ times greater than observed on Saturn. However, one may argue that a planet can host multiple thunderstorms at the same time, so the SED rates are only true for the specific storm and not for the whole planet. Considering an average storm size of 2000 km on Saturn\footnote{Also, similar storm size was observed in December 2010 \citep{Fischer2011b}} \citep{Hurley2012}, the flash density based on the average SED rate of the 2010/2011-storm is $9 \times 10^{-3}$ flashes km$^{-2}$ h$^{-1}$. This flash density is $\sim 4.2 \times 10^{7}$ times smaller than the calculated $3.8 \times 10^5$ flashes km$^{-2}$ h$^{-1}$ on HAT-P-11b.

On Earth, the highest flash density observed, $\sim 0.1$ flashes km$^{-2}$ h$^{-1}$ \citep{Huffines1999}, is produced in thunderstorms within the United States (USA). The 3.87 mJy signal from $\rho_{\rm fl} \approx 3.8 \times 10^5$ flashes km$^{-2}$ h$^{-1}$ on HAT-P-11b, would require a sustained global storm with flash densities $\sim 3.8 \times 10^6$ times greater than observed within the USA. 
 
The obtained average lightning flash density of $3.8 \times 10^5$ flashes km$^{-2}$ h$^{-1}$ seems unlikely to occur based on lightning energetic properties known from the Solar system. The production of much more powerful lightning flashes should be expected on exoplanets different from Solar system planets. The observational implications for direct detection of lightning and for observing its chemical effects on extrasolar planets are discussed in the following sections.

\section{Lightning detection in the optical range}
\label{sec:other}

The emitted power of lightning has been measured in other wavelengths on Jupiter and Saturn. In this section we apply the above-derived flash densities and estimate the emitted optical flux of the lightning storm that could produce the observed radio flux on HAT-P-11b.

\citet[][table 2]{Dyudina2013} lists the survey time (1.9 s), the total optical power ($1.2 \times 10^{10}$ W) and the optical flash rate (5 s$^{-1}$) of the large thunderstorm on Saturn in 2011. Based on this information the average optical power released by a single flash of this Saturnian thunderstorm is $P_{\rm opt,fl} \approx 1.3 \times 10^9$ W. Assuming that flashes on HAT-P-11b produce the same amount of power as Saturnian flashes and using equation (\ref{eq:opt}) we obtain an optical irradiance from a single flash to be $I_{\rm opt,fl} = 1.13 \times 10^{-14}$ Jy.

\begin{equation}
\label{eq:opt}
I_{\rm opt,fl} = \frac{P_{\rm opt,fl}/f_{\rm eff}}{4 \uppi d^2} \times 10^{26},
\end{equation}

\noindent where $P_{\rm opt,fl}$ is the optical power of a single flash and $f_{\rm eff}\approx 6.47 \times 10^{14}$ Hz is the effective frequency of \textit{Cassini}'s blue filter. The total optical irradiance of flashes is obtained from $I_{\rm opt,fl}$ and $n_{\rm tot,fl}$, the total number of flashes hypothetically producing the same radio flux as was observed by \citetalias{Lecav2013}. This optical irradiance is on the order of $10^{-2}$ Jy or brightness of $\sim 13$ mag (B band) as shown in Fig \ref{fig:rflux}, right-hand panel (red). The star HAT-P-11 has apparent B magnitude = 10.66 \citep{Hog2000}, which is $\sim 0.2$ Jy in the B band. The optical emission resulting from lightning, therefore, would be slightly lower than that of the star. 

We carried out the same calculations to determine the planetary and stellar flux ratio, in case lightning on HAT-P-11b emitted a power on the order of the power of super-bolts on Earth, $P_{\rm opt,fl} \approx 10^{12}$ W \citep[][p. 164]{Rakov2003}. The produced optical flux densities and the corresponding magnitude scale are shown in Fig. \ref{fig:rflux}, right-hand panel (blue). The ratio of the planetary lightning flux and the flux of the star (with lightning flux density of $\sim 10$ Jy, Fig. \ref{fig:rflux}, right-hand panel) is $\sim 10^2$. If every single lightning flash on HAT-P-11b would emit $\sim 10^{12}$ W, the optical emission of lightning that would produce the radio emission, would outshine the host star by two orders of magnitude.

\section{Lightning Chemistry}
\label{sec:chem}

We consider the effect of large, Saturn-like lightning storms on HCN chemistry. \citet{Lewis1980} estimated that lightning produces HCN at a rate of $2 \times 10^{-10}$ kg/J. Their model was set up for Jupiter, and is applicable for any hydrogen dominated atmosphere with roughly solar composition. \citet{Farrell2007} estimated the dissipative energy of a single lightning flash on Saturn, with peak frequency, $f_0 = 10$ kHz, and spectral roll-off, $n = 3.5$, to be:

\begin{equation}
E_d \approx 260 \, {\rm J} \; \Bigg(\dfrac{f_{\rm Sat}}{f_0}\Bigg)^{\!\!n},
\end{equation}

\noindent where $f_{\rm Sat} = 10$ MHz is the frequency at which the lightning on Saturn was observed. We can multiply the dissipative energy by the flash density of 1 flash km$^{-2}$ h$^{-1}$. Multiplying the lightning energy density by the production rate of HCN, we estimate that $5 \times 10^{-7}$ kg m$^{-2}$ s$^{-1}$ of HCN is produced, of the order of $10^9$ greater than the estimate of \citet{Lewis1980} for Jupiter. Accepting the energetics arguments from \citet{Lewis1980}, the resulting HCN will achieve a volume mixing ratio of $\sim 10^{-6}$ within the mbar regime of the atmosphere. \citet{Moses2013} found that similar mixing ratios (their fig. 11) should have significant observational consequences in the \textit{L} ($3.0-4.0 \mu$m) and \textit{N} ($7.5-14.5 \mu$m) IR bands, which they show in their fig. 16 comparing their model spectra both with and without HCN.

\begin{figure}
\includegraphics[width=\linewidth]{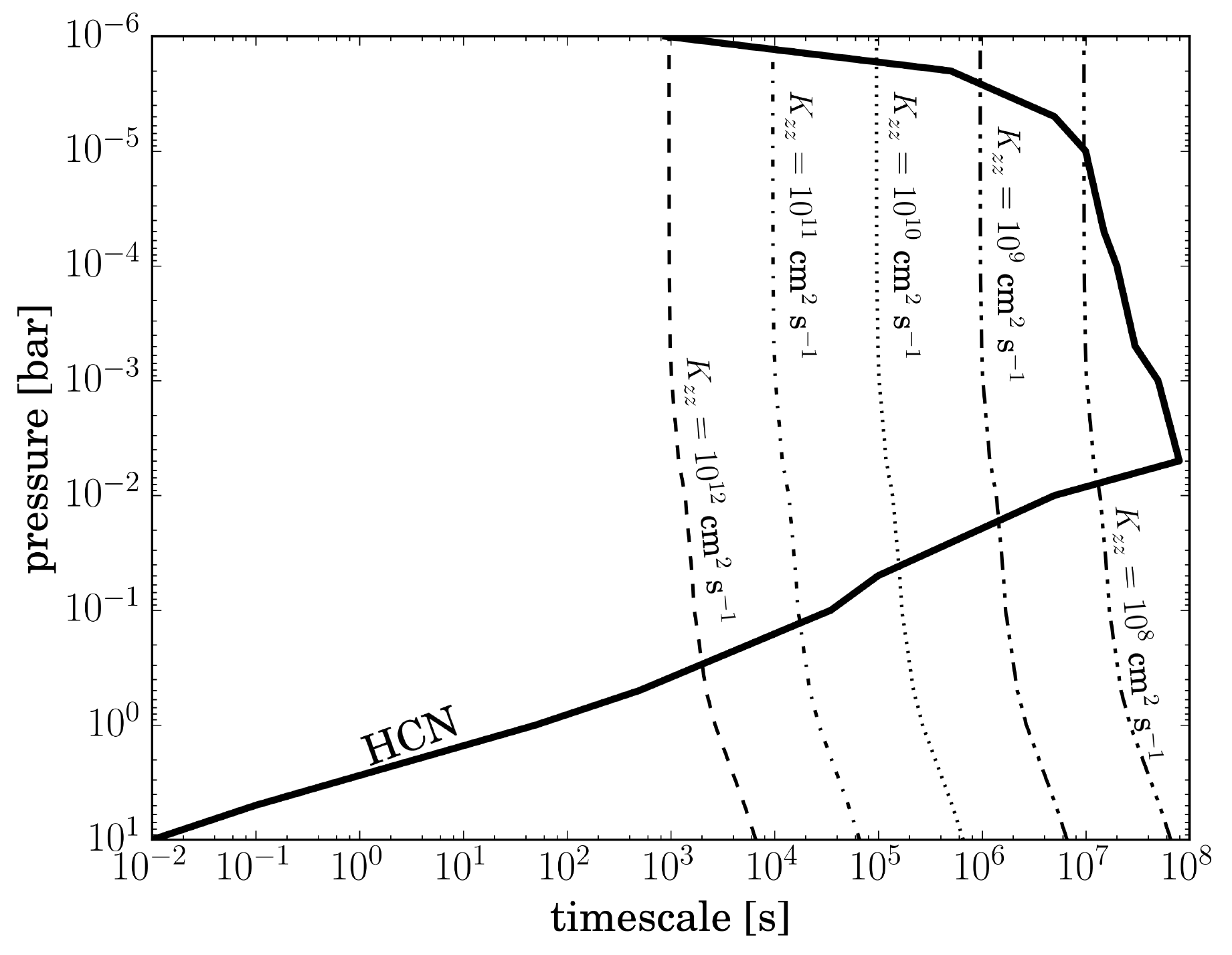}
\caption{The chemical lifetime of HCN [s] plotted vs. pressure [bar], along with various dynamical timescales with eddy diffusion coefficients ranging $K_{zz} = 10^8 \,\ldots\, 10^{12}$ cm$^2$ s$^{-1}$.}
\label{fig:hcn}
\end{figure}

In order to estimate the chemical timescale\footnote{The amount of time necessary for the atmospheric abundance of HCN to achieve thermochemical equilibrium.} for HCN on HAT-P-11b, we develop a semi-analytical pressure$-$temperature profile appropriate for the object using the method of \citet{Hansen2008}. We use for this profile the parameters for HAT-P-11b (mass, radius, distance from host star) given by \citet{Bakos2010} and \citet{Lopez2014}, and the stellar temperature from \citet{Bakos2010}. To determine the XUV flux impinging on the atmosphere, we take a spectrum appropriate for a K4-type star, the X-Exospheres synthetic spectrum for HD 111232 \citep{Sanz-Forcada2011}. We assume that the atmosphere is hydrogen rich, and the atmospheric gas of HAT-P-11b is at roughly solar metallicity with respect to C, N and O, i.e. that the primordial concentrations of these elements in HAT-P-11b is solar and that there is no elemental depletion into clouds. We calculate the atmospheric chemistry using our semi-analytic temperature profile and synthetic XUV flux, with the STAND2015 chemical network and the ARGO diffusion-photochemistry model \citep{Rimmer2015}. We then examine a variety of locations in the atmosphere, injecting HCN at a mixing ratio of $10^{-6}$, and evolving the atmospheric chemistry in time to determine the chemical timescale for HCN as a function of pressure, shown in Fig. \ref{fig:hcn}. The dynamical timescale for vertical mixing is overlaid on the top of this plot, assuming a range of constant eddy diffusion coefficients.

The chemical timescale for HCN ranges from 100 milliseconds at the bottom of our model atmosphere (10 bar), to 2.5 yr at 5 mbar (Fig. \ref{fig:hcn}). At pressures less than 5 mbar, the timescale for HCN slowly drops down to about 4 months at 10 $\mu$bar, and then drops precipitously at lower pressures, until at 1 $\mu$bar it achieves a timescale of about 30 minutes (Fig. \ref{fig:hcn}). These results can be compared to the dynamical timescale of the atmosphere, represented approximately by the eddy diffusion coefficient \citep[see][for details]{Lee2015}. If the chemical timescale for HCN is smaller than the dynamical timescale, then the HCN would be destroyed before it is transported higher into the atmosphere. If the chemical timescale for HCN is larger than the dynamical timescale, then the HCN will survive long enough to reach other parts of the atmosphere, where it will survive longer. Assuming that lightning takes places on HAT-P-11b at pressures of $\lesssim 0.1$ bar, the HCN will survive long enough to be transported into the mbar regime, where it will survive for $2-3$ yr before being chemically destroyed. If, on the other hand, HCN is formed much below the 0.1 bar level, at pressures of $\gtrsim 1$ bar, the chemical timescale is too short for the HCN to escape, and it will be rapidly destroyed before it could be observed.

It is informative to examine how HCN is destroyed. In the lower atmosphere, HCN is primarily destroyed by reacting with atomic hydrogen: ${\rm HCN + H} \rightarrow {\rm CN + H_2}$. The CN will quickly react with H$_2$ to reform HCN, but at the same time, the thermally dissociated products of water, the hydroxyl radical, enters into a rapid three-body reaction with CN to form HCNO, which reacts with H to form NO and CH$_2^1$. This process quickly becomes inefficient higher in the atmosphere, both because the termolecular reaction to form HCNO becomes less efficient and because less atomic hydrogen is available. However, in the upper atmosphere, photochemistry begins to set the chemical time-scale for HCN, primarily via the reaction: ${\rm HCN + O} \rightarrow {\rm NCO + H}$. The atomic oxygen is liberated by photolysis of the hydroxyl radical, which itself results from reactions between water and the photochemical products of methane.

\section{Summary}
\label{sec:summary}

We present an interpretation of the radio observations of HAT-P-11b made by \citetalias{Lecav2013} under the assumption that these transient radio signals are real and were caused by lightning on HAT-P-11b. We determined that the flash density of $3.8 \times 10^5$ flashes km$^{-2}$ h$^{-1}$ over the hemisphere of the planet would be necessary to explain the average radio signal, $3.8 \times 10^6$ times greater than the rate of lightning storms observed in dense storms within the United States and many orders of magnitude beyond the storms observed on Saturn. We also examine the optical emission such a storm would generate, as well as the impact of this storm on the atmospheric chemistry, assuming a hydrogen-rich atmosphere.

In summary, we find that

\begin{enumerate}
\item the radio emission of a few mJy at 150 MHz, at the distance of HAT-P-11b, requires unrealistically high flash densities if this lightning is like in the Solar system. Therefore, lightning produced radio emission most probably cannot be observed by current radio telescopes at frequencies of 150 MHz or higher.

\item The optical counterpart of the enormous lightning storm would be as bright as the host star itself.

\item The amount of HCN produced by lightning at pressures $\leq 0.1$ bar, in a hydrogen-rich atmosphere of an irradiated exoplanet with strong winds, may in some cases yield detectable quantities that linger in the atmosphere for $2-3$ yr after the advent of the lightning storm. If the lightning occurs much deeper in the atmosphere, the HCN will react away before it can diffuse into the upper atmosphere, and will probably not be observable.
\end{enumerate}

Our results show that the radio emission on HAT-P-11b is unlikely to be caused by lightning, if lightning properties similar to the Solar system ones are assumed. However, it shows a new interpretation that could be applied to high frequency (up to $\sim30-50$ MHz) radio observations, where it is more probable to observe lightning, because of its radiating properties (equation (\ref{eq:1})) Our calculations can be also applied to determine the minimum storm size detectable within an exoplanetary atmosphere using current or future radio instruments. Our recommendation to observers who detect radio emission in the frequency range of a few tens of MHz, especially if it is unpolarized, from an exoplanet would be to follow up these observations with infrared observations made in the \textit{L} and \textit{N} bands when possible, in order to look for HCN emission, which should be observable for $2-3$ yr if lightning occurs around the 0.1 bar level of an atmosphere with reasonably large vertical convective velocities. If HCN is detected at that time, and if both the radio emission and the HCN turn out to be transient, this would be strong evidence for lightning on an exoplanet.

\section*{Acknowledgements}
\small 
We thank Zach Cano, Brice-Olivier Demory, Aurora Sicilia-Aguilar and Alain Lecavelier des Etangs for useful discussions. We thank Philippe Zarka for pointing out an error in our equations. We highlight financial support of the European Community under the FP7 by an ERC starting grant number 257431.

\bibliographystyle{mnras}
\bibliography{bib_hatp}

\begin{thebibliography}{}
\makeatletter
\relax
\def\mn@urlcharsother{\let\do\@makeother \do\$\do\&\do\#\do\^\do\_\do\%\do\~}
\def\mn@doi{\begingroup\mn@urlcharsother \@ifnextchar [ {\mn@doi@}
  {\mn@doi@[]}}
\def\mn@doi@[#1]#2{\def\@tempa{#1}\ifx\@tempa\@empty \href
  {http://dx.doi.org/#2} {doi:#2}\else \href {http://dx.doi.org/#2} {#1}\fi
  \endgroup}
\def\mn@eprint#1#2{\mn@eprint@#1:#2::\@nil}
\def\mn@eprint@arXiv#1{\href {http://arxiv.org/abs/#1} {{\tt arXiv:#1}}}
\def\mn@eprint@dblp#1{\href {http://dblp.uni-trier.de/rec/bibtex/#1.xml}
  {dblp:#1}}
\def\mn@eprint@#1:#2:#3:#4\@nil{\def\@tempa {#1}\def\@tempb {#2}\def\@tempc
  {#3}\ifx \@tempc \@empty \let \@tempc \@tempb \let \@tempb \@tempa \fi \ifx
  \@tempb \@empty \def\@tempb {arXiv}\fi \@ifundefined
  {mn@eprint@\@tempb}{\@tempb:\@tempc}{\expandafter \expandafter \csname
  mn@eprint@\@tempb\endcsname \expandafter{\@tempc}}}

\bibitem[\protect\citeauthoryear{{Aplin}}{{Aplin}}{2013}]{Aplin2013}
{Aplin} K.~L.,  2013, {Electrifying Atmospheres: Charging, Ionisation and
  Lightning in the Solar System and Beyond}.
SpringerBriefs in Astronomy, Springer Science+Business Media Dordrecht,
  \mn@doi{10.1007/978-94-007-6633-4}

\bibitem[\protect\citeauthoryear{{Bailey}, {Helling}, {Hodos{\'a}n}, {Bilger}
  \& {Stark}}{{Bailey} et~al.}{2014}]{Bailey2014}
{Bailey} R.~L.,  {Helling} C.,  {Hodos{\'a}n} G.,  {Bilger} C.,   {Stark}
  C.~R.,  2014, \mn@doi [\apj] {10.1088/0004-637X/784/1/43}, \href
  {http://adsabs.harvard.edu/abs/2014ApJ...784...43B} {784, 43}

\bibitem[\protect\citeauthoryear{{Bakos} et~al.,}{{Bakos}
  et~al.}{2010}]{Bakos2010}
{Bakos} G.~{\'A}.,  et~al., 2010, \mn@doi [\apj]
  {10.1088/0004-637X/710/2/1724}, \href
  {http://adsabs.harvard.edu/abs/2010ApJ...710.1724B} {710, 1724}

\bibitem[\protect\citeauthoryear{{Dyudina}, {Ingersoll}, {Ewald}, {Porco},
  {Fischer}  \& {Yair}}{{Dyudina} et~al.}{2013}]{Dyudina2013}
{Dyudina} U.~A.,  {Ingersoll} A.~P.,  {Ewald} S.~P.,  {Porco} C.~C.,  {Fischer}
  G.,   {Yair} Y.,  2013, \mn@doi [\icarus] {10.1016/j.icarus.2013.07.013},
  \href {http://adsabs.harvard.edu/abs/2013Icar..226.1020D} {226, 1020}

\bibitem[\protect\citeauthoryear{{Farrell}, {Kaiser}, {Fischer}, {Zarka},
  {Kurth}  \& {Gurnett}}{{Farrell} et~al.}{2007}]{Farrell2007}
{Farrell} W.~M.,  {Kaiser} M.~L.,  {Fischer} G.,  {Zarka} P.,  {Kurth} W.~S.,
  {Gurnett} D.~A.,  2007, \mn@doi [\grl] {10.1029/2006GL028841}, \href
  {http://adsabs.harvard.edu/abs/2007GeoRL..34.6202F} {34, 6202}

\bibitem[\protect\citeauthoryear{{Fischer} et~al.,}{{Fischer}
  et~al.}{2006}]{Fischer2006}
{Fischer} G.,  et~al., 2006, \mn@doi [\icarus] {10.1016/j.icarus.2006.02.010},
  \href {http://adsabs.harvard.edu/abs/2006Icar..183..135F} {183, 135}

\bibitem[\protect\citeauthoryear{{Fischer}, {Kurth}, {Dyudina}, {Kaiser},
  {Zarka}, {Lecacheux}, {Ingersoll}  \& {Gurnett}}{{Fischer}
  et~al.}{2007}]{Fischer2007}
{Fischer} G.,  {Kurth} W.~S.,  {Dyudina} U.~A.,  {Kaiser} M.~L.,  {Zarka} P.,
  {Lecacheux} A.,  {Ingersoll} A.~P.,   {Gurnett} D.~A.,  2007, \mn@doi
  [\icarus] {10.1016/j.icarus.2007.04.002}, \href
  {http://adsabs.harvard.edu/abs/2007Icar..190..528F} {190, 528}

\bibitem[\protect\citeauthoryear{{Fischer} et~al.,}{{Fischer}
  et~al.}{2011}]{Fischer2011b}
{Fischer} G.,  et~al., 2011, \mn@doi [\nat] {10.1038/nature10205}, \href
  {http://adsabs.harvard.edu/abs/2011Natur.475...75F} {475, 75}

\bibitem[\protect\citeauthoryear{{Fraine} et~al.,}{{Fraine}
  et~al.}{2014}]{Fraine2014}
{Fraine} J.,  et~al., 2014, \mn@doi [\nat] {10.1038/nature13785}, \href
  {http://adsabs.harvard.edu/abs/2014Natur.513..526F} {513, 526}

\bibitem[\protect\citeauthoryear{{Grie{\ss}meier}, {Zarka}  \&
  {Spreeuw}}{{Grie{\ss}meier} et~al.}{2007}]{Griessmeier2007}
{Grie{\ss}meier} J.-M.,  {Zarka} P.,   {Spreeuw} H.,  2007, \mn@doi [\aap]
  {10.1051/0004-6361:20077397}, \href
  {http://adsabs.harvard.edu/abs/2007A%26A...475..359G} {475, 359}

\bibitem[\protect\citeauthoryear{{Gurnett}, {Shaw}, {Anderson}  \&
  {Kurth}}{{Gurnett} et~al.}{1979}]{Gurnett1979}
{Gurnett} D.~A.,  {Shaw} R.~R.,  {Anderson} R.~R.,   {Kurth} W.~S.,  1979,
  \mn@doi [\grl] {10.1029/GL006i006p00511}, \href
  {http://adsabs.harvard.edu/abs/1979GeoRL...6..511G} {6, 511}

\bibitem[\protect\citeauthoryear{{Hansen}}{{Hansen}}{2008}]{Hansen2008}
{Hansen} B.~M.~S.,  2008, \mn@doi [\apjs] {10.1086/591964}, \href
  {http://adsabs.harvard.edu/abs/2008ApJS..179..484H} {179, 484}

\bibitem[\protect\citeauthoryear{{Helling}, {Jardine}, {Stark}  \&
  {Diver}}{{Helling} et~al.}{2013}]{Helling2013}
{Helling} C.,  {Jardine} M.,  {Stark} C.,   {Diver} D.,  2013, \mn@doi [\apj]
  {10.1088/0004-637X/767/2/136}, \href
  {http://esoads.eso.org/abs/2013ApJ...767..136H} {767, 136}

\bibitem[\protect\citeauthoryear{{Helling} et~al.,}{{Helling}
  et~al.}{2016}]{Helling2016}
{Helling} C.,  et~al., 2016, preprint, \href
  {http://adsabs.harvard.edu/abs/2016arXiv160304022H} {} (\mn@eprint {arXiv}
  {1603.04022})

\bibitem[\protect\citeauthoryear{{Heng} \& {Showman}}{{Heng} \&
  {Showman}}{2015}]{Heng2015}
{Heng} K.,  {Showman} A.~P.,  2015, \mn@doi [Annual Review of Earth and
  Planetary Sciences] {10.1146/annurev-earth-060614-105146}, \href
  {http://adsabs.harvard.edu/abs/2015AREPS..43..509H} {43, 509}

\bibitem[\protect\citeauthoryear{{Hodos\'an}, {Helling}, {Asensio-Torres},
  {Vorgul}  \& {Rimmer}}{{Hodos\'an} et~al.}{2016}]{Hodosan2016a}
{Hodos\'an} G.,  {Helling} C.,  {Asensio-Torres} R.,  {Vorgul} I.,   {Rimmer}
  P.,  2016, \mnras, p. {submitted}

\bibitem[\protect\citeauthoryear{{H{\o}g} et~al.,}{{H{\o}g}
  et~al.}{2000}]{Hog2000}
{H{\o}g} E.,  et~al., 2000, \aap, \href
  {http://cdsads.u-strasbg.fr/abs/2000A%26A...355L..27H} {355, L27}

\bibitem[\protect\citeauthoryear{{Huffines} \& {Orville}}{{Huffines} \&
  {Orville}}{1999}]{Huffines1999}
{Huffines} G.~R.,  {Orville} R.~E.,  1999, \mn@doi [Journal of Applied
  Meteorology] {10.1175/1520-0450(1999)038<1013:LGFDAT>2.0.CO;2}, \href
  {http://adsabs.harvard.edu/abs/1999JApMe..38.1013H} {38, 1013}

\bibitem[\protect\citeauthoryear{{Hurley}, {Irwin}, {Fletcher}, {Moses},
  {Hesman}, {Sinclair}  \& {Merlet}}{{Hurley} et~al.}{2012}]{Hurley2012}
{Hurley} J.,  {Irwin} P.~G.~J.,  {Fletcher} L.~N.,  {Moses} J.~I.,  {Hesman}
  B.,  {Sinclair} J.,   {Merlet} C.,  2012, \mn@doi [\planss]
  {10.1016/j.pss.2011.12.026}, \href
  {http://adsabs.harvard.edu/abs/2012P%26SS...65...21H} {65, 21}

\bibitem[\protect\citeauthoryear{{Lecavelier des Etangs}, {Sirothia},
  {Gopal-Krishna}  \& {Zarka}}{{Lecavelier des Etangs}
  et~al.}{2013}]{Lecav2013}
{Lecavelier des Etangs} A.,  {Sirothia} S.~K.,  {Gopal-Krishna}  {Zarka} P.,
  2013, \mn@doi [\aap] {10.1051/0004-6361/201219789}, \href
  {http://adsabs.harvard.edu/abs/2013A%26A...552A..65L} {552, A65}

\bibitem[\protect\citeauthoryear{{Lee}, {Helling}, {Dobbs-Dixon}  \&
  {Juncher}}{{Lee} et~al.}{2015}]{Lee2015}
{Lee} G.,  {Helling} C.,  {Dobbs-Dixon} I.,   {Juncher} D.,  2015, \mn@doi
  [\aap] {10.1051/0004-6361/201525982}, \href
  {http://adsabs.harvard.edu/abs/2015A%26A...580A..12L} {580, A12}

\bibitem[\protect\citeauthoryear{{Lewis}}{{Lewis}}{1980}]{Lewis1980}
{Lewis} J.~S.,  1980, \mn@doi [\icarus] {10.1016/0019-1035(80)90090-1}, \href
  {http://adsabs.harvard.edu/abs/1980Icar...43...85L} {43, 85}

\bibitem[\protect\citeauthoryear{{Line} \& {Parmentier}}{{Line} \&
  {Parmentier}}{2016}]{Line2016}
{Line} M.~R.,  {Parmentier} V.,  2016, \mn@doi [\apj]
  {10.3847/0004-637X/820/1/78}, \href
  {http://adsabs.harvard.edu/abs/2016ApJ...820...78L} {820, 78}

\bibitem[\protect\citeauthoryear{{Lopez} \& {Fortney}}{{Lopez} \&
  {Fortney}}{2014}]{Lopez2014}
{Lopez} E.~D.,  {Fortney} J.~J.,  2014, \mn@doi [\apj]
  {10.1088/0004-637X/792/1/1}, \href
  {http://adsabs.harvard.edu/abs/2014ApJ...792....1L} {792, 1}

\bibitem[\protect\citeauthoryear{{Mayor} \& {Queloz}}{{Mayor} \&
  {Queloz}}{1995}]{Mayor1995}
{Mayor} M.,  {Queloz} D.,  1995, \mn@doi [\nat] {10.1038/378355a0}, \href
  {http://adsabs.harvard.edu/abs/1995Natur.378..355M} {378, 355}

\bibitem[\protect\citeauthoryear{{Moses}, {Madhusudhan}, {Visscher}  \&
  {Freedman}}{{Moses} et~al.}{2013}]{Moses2013}
{Moses} J.~I.,  {Madhusudhan} N.,  {Visscher} C.,   {Freedman} R.~S.,  2013,
  \mn@doi [\apj] {10.1088/0004-637X/763/1/25}, \href
  {http://adsabs.harvard.edu/abs/2013ApJ...763...25M} {763, 25}

\bibitem[\protect\citeauthoryear{{Rakov} \& {Uman}}{{Rakov} \&
  {Uman}}{2003}]{Rakov2003}
{Rakov} V.~A.,  {Uman} M.~A.,  2003, {Lightning}.
Cambridge University Press, Cambridge, UK

\bibitem[\protect\citeauthoryear{{Rimmer} \& {Helling}}{{Rimmer} \&
  {Helling}}{2015}]{Rimmer2015}
{Rimmer} P.~B.,  {Helling} C.,  2015, preprint, \href
  {http://adsabs.harvard.edu/abs/2015arXiv151007052R} {} (\mn@eprint {arXiv}
  {1510.07052})

\bibitem[\protect\citeauthoryear{{Russell}, {Zhang}  \& {Wei}}{{Russell}
  et~al.}{2008}]{Russell2008}
{Russell} C.~T.,  {Zhang} T.~L.,   {Wei} H.~Y.,  2008, \mn@doi [Journal of
  Geophysical Research (Space Physics)] {10.1029/2008JE003137}, \href
  {http://adsabs.harvard.edu/abs/2008JGRA..113.0B05R} {113, 0}

\bibitem[\protect\citeauthoryear{{Sanz-Forcada}, {Micela}, {Ribas}, {Pollock},
  {Eiroa}, {Velasco}, {Solano}  \& {Garc{\'{\i}}a-{\'A}lvarez}}{{Sanz-Forcada}
  et~al.}{2011}]{Sanz-Forcada2011}
{Sanz-Forcada} J.,  {Micela} G.,  {Ribas} I.,  {Pollock} A.~M.~T.,  {Eiroa} C.,
   {Velasco} A.,  {Solano} E.,   {Garc{\'{\i}}a-{\'A}lvarez} D.,  2011, \mn@doi
  [\aap] {10.1051/0004-6361/201116594}, \href
  {http://adsabs.harvard.edu/abs/2011A%26A...532A...6S} {532, A6}

\bibitem[\protect\citeauthoryear{{Shao} \& {Jacobson}}{{Shao} \&
  {Jacobson}}{2002}]{Shao2002}
{Shao} X.-M.,  {Jacobson} A.~R.,  2002, \mn@doi [Journal of Geophysical
  Research (Atmospheres)] {10.1029/2001JD001018}, \href
  {http://adsabs.harvard.edu/abs/2002JGRD..107.4430S} {107, 4430}

\bibitem[\protect\citeauthoryear{{Vorgul} \& {Helling}}{{Vorgul} \&
  {Helling}}{2016}]{Vorgul2016}
{Vorgul} I.,  {Helling} C.,  2016, \mn@doi [\mnras] {10.1093/mnras/stw234},
  \href {http://adsabs.harvard.edu/abs/2016MNRAS.458.1041V} {458, 1041}

\bibitem[\protect\citeauthoryear{{Weibel}}{{Weibel}}{1959}]{Weibel1959}
{Weibel} E.~S.,  1959, \mn@doi [Physical Review Letters]
  {10.1103/PhysRevLett.2.83}, \href
  {http://adsabs.harvard.edu/abs/1959PhRvL...2...83W} {2, 83}

\bibitem[\protect\citeauthoryear{{Williams} \& {Berger}}{{Williams} \&
  {Berger}}{2015}]{Williams2015}
{Williams} P.~K.~G.,  {Berger} E.,  2015, \mn@doi [\apj]
  {10.1088/0004-637X/808/2/189}, \href
  {http://adsabs.harvard.edu/abs/2015ApJ...808..189W} {808, 189}

\bibitem[\protect\citeauthoryear{{Wolszczan} \& {Frail}}{{Wolszczan} \&
  {Frail}}{1992}]{Wolszczan1992}
{Wolszczan} A.,  {Frail} D.~A.,  1992, \mn@doi [\nat] {10.1038/355145a0}, \href
  {http://adsabs.harvard.edu/abs/1992Natur.355..145W} {355, 145}

\bibitem[\protect\citeauthoryear{{Zarka} \& {Pedersen}}{{Zarka} \&
  {Pedersen}}{1986}]{Zarka1986}
{Zarka} P.,  {Pedersen} B.~M.,  1986, \mn@doi [\nat] {10.1038/323605a0}, \href
  {http://adsabs.harvard.edu/abs/1986Natur.323..605Z} {323, 605}

\bibitem[\protect\citeauthoryear{{Zarka}, {Farrell}, {Kaiser}, {Blanc}  \&
  {Kurth}}{{Zarka} et~al.}{2004}]{Zarka2004}
{Zarka} P.,  {Farrell} W.~M.,  {Kaiser} M.~L.,  {Blanc} E.,   {Kurth} W.~S.,
  2004, \mn@doi [\planss] {10.1016/j.pss.2004.09.011}, \href
  {http://adsabs.harvard.edu/abs/2004P%26SS...52.1435Z} {52, 1435}

\makeatother
\end{thebibliography}

\appendix

\bsp	
\label{lastpage}
\end{document}